\documentclass[journal=nalefd,manuscript=article]{achemso}

\usepackage{amssymb} 
\usepackage{amsmath} 
\usepackage[utf8]{inputenc} 
\usepackage[sort&compress,numbers]{natbib}
\pdfoutput=1 

\usepackage{siunitx}  
\usepackage{graphicx}
\usepackage[export]{adjustbox}
\usepackage{hyperref}
\usepackage{float}

\author{Niklas Müller}
\altaffiliation{Both authors contributed equally}
\author{Vincent Hock}
\altaffiliation{Both authors contributed equally}
\author{Holger Koch}
\author{Nora Bach}
\author{Christopher Rathje}
\author{Sascha Schäfer}
\email{sascha.schaefer@uni-oldenburg.de}
\affiliation[University]
{Institute of Physics, University of Oldenburg, 26129 Oldenburg, Germany}

\title{Broadband coupling of fast electrons to high-Q whispering-gallery mode resonators}


\begin{document}

	\begin{abstract}
		Transmission electron microscopy is an excellent experimental tool to study the interaction of free electrons with nanoscale light fields.  However, up to now, applying electron microscopy to quantum optical investigations was hampered by the lack of experimental platforms which allow a strong coupling between fast electrons and high-quality resonators. Here, as a first step, we demonstrate the broad-band excitation of optical whispering-gallery modes in silica microresonators by fast electrons. In the emitted coherent cathodoluminescence spectrum, a comb of equidistant peaks is observed, resulting in cavity quality factors larger than 700. These results enable the study of quantum optical phenomena in electron microscopy with potential applications in quantum electron-light metrology.
	\end{abstract}

	\section{}
	The inelastic interaction of fast electrons with nanostructures in transmission electron microscopy (TEM) has provided a detailed picture of nano-optical properties, due to the dependence of the spontaneous electron energy loss (EEL) probabilities on the local density of photonic states \cite{de2010optical,cai2009efficient,ogut2012toroidal,von2013isolated,schroder2015real,talebi2015excitation,yalunin2016theory}. The coupling of local photonic states to far-field light and the spectral characteristics of nanoscale photon emitters can be mapped by cathodoluminescence (CL) spectroscopy \cite{de2010optical,myroshnychenko2012plasmon}.  Additionally, harnessing the enhanced transition probability provided by stimulated relative to spontaneous processes, the interaction of fast electrons with laser-driven intense nanoscale optical near-fields has recently enabled novel techniques both for the characterization of optical near-fields \cite{de2008electron,barwick2009photon,park2010photon,garcia2010multiphoton} as well as for the amplitude and phase-tailoring of free-electron wavefunctions \cite{feist2015quantum,echternkamp2016ramsey,kfir2019controlling,schwartz2019laser,vanacore2019ultrafast,feist2020high}. The precise control capabilities of coherent optical fields on electron wavefunctions even enabled the generation of attosecond electron pulse trains \cite{priebe2017attosecond,morimoto2018diffraction,kozak2018ponderomotive}. 
	
	Despite the rapid experimental progress recently made regarding the interaction of electrons with coherent light, only a limited number of studies have addressed the coupling of non-classical (quantum optical) light states with free electrons. As one of the first examples in this direction, it was demonstrated that the intensity autocorrelation of the emitted CL light from defect centers in hexagonal boron nitride exhibits photon-bunching \cite{meuret2015photon}. 
	
	Historically, progress in quantum optics heavily relied on the development of high-quality (high-Q) optical cavities coupled to few-level quantum systems and has resulted in the triving field of cavity quantum electrodynamics \cite{berman1994cavity,kimble1998strong,raimond2001manipulating,mabuchi2002cavity,vahala2003optical,spillane2005ultrahigh,walther2006cavity}. As a prominent example, optical whispering-gallery modes (WGM) \cite{matsko2006optical,ilchenko2006optical} within dielectric microresonators \cite{vahala2003optical,kavokin2017microcavities} exhibit quality factors exceeding $10^9$ with mode lifetimes larger than \SI{1} {\micro\second} \cite{collot1993very,gorodetsky1996ultimate}. Frequency combs in such resonators \cite{kippenberg2011microresonator,del2007optical} are already finding wide applications for high-resolution spectroscopy \cite{suh2016microresonator,bernhardt2010cavity}, the development of precise optical clocks and optical synthesizers \cite{diddams2001optical,spencer2018optical}, for absolute distance measurement \cite{coddington2009rapid} and for the sensitive exoplanet search in astronomy \cite{suh2019searching}.
	
	A similar line of development in transmission electron microscopy was previously hampered by the lack of available experimental means to couple free electrons with high-Q optical cavities, despite recent progress in the coupling of free electrons to low-Q nanoscale dielectric resonators \cite{hyun2008measuring,hyun2010relativistic,talebi2019merging} and the stimulated interaction with photonic-crystal cavity modes \cite{wang2020coherent}. 
	
	Here, we demonstrate the electron-induced excitation of high-Q whispering-gallery modes in silica microfibres in a TEM. Fibre-guided coherent cathodoluminescence is observed and the CL spectra consist of an octave-spanning frequency comb with narrow-bandwidth peaks. In agreement with numerical simulations, the peaks within the comb exhibit a spacing inversely scaling with the fibre circumference and show a peak width down to below 3.5 meV. Our findings establish silica microfibres as a novel experimental platform for the coherent electron-light coupling in dielectrics, possibly enabling the investigation of strong-coupling and quantum entanglement phenomena in electron microscopy.

	In the experiment, we focus a 200-keV electron beam (focal diameter: 10 nm) close to the surface of a silica microfibre tip, as sketched in Fig. 1 a,b. The electrons travel at a speed of about $0.7\,c$ ($c$: speed of light) and generate an evanescent electromagnetic field which induces a transient electric polarization in the silica microfibre.  Efficient coupling requires nanoscale distances between the electron and the dielectric, so that the moving charge interacts with the dielectric only during a short period time (about 1 fs in the present case).  
	The induced femtosecond polarization pulse (with field orientation perpendicular to the fibre surface, corresponding to TM-modes) is partially guided within the microfibre and detected by a low-noise spectrometer at the non-tapered fibre output. A typical coherent CL spectrum recorded at a distance of $d$ = \SI{130} {\micro\metre} from the tip apex is shown in Fig. 1c. The spectrum extends from a photon energy of 1.4 to 3~eV and consists of a series of equidistant primary peaks with a peak spacing of 35~meV (Fig. 1c, inset). Depending on the spectral region, additional substructures are visible between the primary peaks. Within the observed frequency range, a total of 1.1 $\cdot$ $10^6$ photons/s are detected (intensity calibration: see methods). Considering an incident electron beam current of 680 pA, spectral CL efficiencies $\Gamma_{\mathrm{CL}}(E)$ in the range of $10^{-4}$/eV and an overall electron-photon conversion efficiency of 0.02\% are obtained. 
	
	We note that in addition to the light generated by electron induced polarizations, i.e. coherent CL, direct excitation of defect centers in silica can occur, in particular when the electron beam is transmitted through the fibre material. In this case, cathodoluminescence at specific defect transitions of silica (around 1.9 and 2.7 eV \cite{goldberg1997cathodoluminescence}) is observed at intensities about 10-100 times larger than the observed coherent CL. Small contributions of inchoherent CL are also visible in Fig. 1c, resulting in a slight variation of the spectral baseline.  Due to the nanosecond to microsecond lifetimes of excited defect states, light emitted from defect centers has no phase relation to the arrival time of the exciting electron beam, unlike for the case of coherent CL.

	\begin{figure*}
		\includegraphics[scale=1]{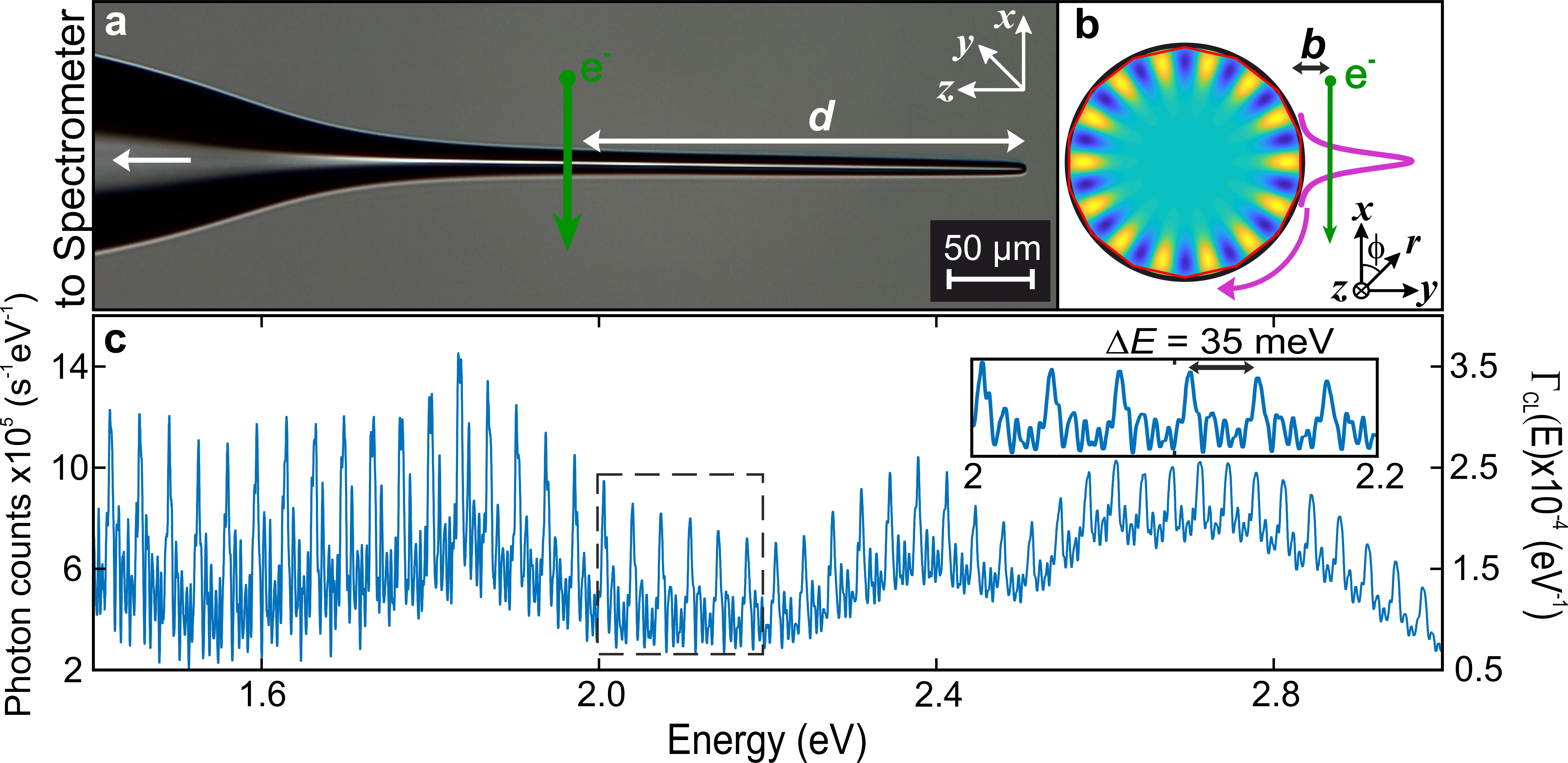}
		\caption{Coherent cathodoluminescence from microfibre tips. \textbf{(a,b)} Schematics of the experimental setup. A nano-focussed 200-keV electron beam passes a fibre tip close to its surface at a distance $d$ from the fibre apex. The evanescent electric field of the electron (impact parameter $b$) induces a femtosecond electric polarization in the fibre, which can be decomposed into optical whispering-gallery modes (WGM) of the fibre taper. Background image in (a): optical micrograph of the tapered silica fibre. As one example, a WGM for $\nu$ = 14 is sketched in (b).  \textbf{(c)} Broadband coherent CL spectrum recorded at an electron beam current of $I$ = \SI{680} pA exhibiting a comb spacing of $\Delta E$ = \SI{35} meV (impact parameter $b~\approx$ 25 nm, $d$ = \SI{130}{\micro\metre}, fibre radius of $\rho$ = \SI{4.5}{\micro\metre}). Inset: Close-up of a spectral region around 2-2.2 eV. $\Gamma_{\mathrm{CL}}$: inelastic scattering efficiency.}
		\label{figure1}
	\end{figure*}
	
	The comb-like spectrum formed by the primary peaks can be intuitively understood in a simplified physical picture, in which an electron-induced light ray is guided by total internal reflection around the circumference of the fibre. Depending on the optical path length during one round-trip, interference results in a resonant enhancement at specific photon energies. For a resonator with refractive index $n$ and radius $\rho$, constructive interference is obtained for a vacuum wavelength $\lambda$ equal to $2\pi \rho \cdot n/ \nu$ where $\nu$ is an integer number. Using the local fibre radius of \SI{4.6} {\micro\metre} at the probe position, a peak spacing of about \SI{30} meV is expected, in reasonable agreement with the experiment. Furthermore, the narrow line width of the primary peaks is indicative of multiple-beam interference, demonstrating that the light fields of multiple round-trips in the ring-type cavity are superposed.
	
	In optical wave theory, the prominent resonant states in such a geometry are commonly known as whispering-gallery modes \cite{matsko2006optical,ilchenko2006optical}, for which the light is confined within a dielectric medium close to the surface with an evanescent component extending into the vacuum. Due to the small opening angle of the fibre tip ($\approx$  \SI{0.3}{\degree}), the fibre can be locally described as being approximately cylindrical. In this case, the resonant modes can be factorized into $\Phi_{\nu,l}(r,\phi)\exp(-i\beta z)$. The transverse modal field $\Phi_{\nu,l}$ is labeled by the azimuthal and radial mode orders $\nu$ and $l$, respectively, and the propagation component along the fibre axis is characterized by the wavenumber $\beta$. An exemplary modal field for $\nu=14$ is sketched in Fig. 1b. The femtosecond polarization pulse induced by the passing electron can be considered as the superposition of a larger number of whispering-gallery modes with a defined relative phase relation.
	We note, that since the spectrometer is connected to the end of the fibre, we only detect whispering-gallery modes which have a significant propagation constant along the fibre axis.  
	
	In order to further investigate how the resonant spectral features depend on the local fibre geometry, we recorded CL spectra (within a selected photon energy window) with the electron beam positioned at various distances $d$, ranging from \SI{100} to \SI{250} {\micro\metre} from the fibre apex. Within this interval, the fibre radius increases from \SI{4.3} to \SI{5.3} {\micro\metre}. For all spectra, the distance between the electron beam and the fibre surface was maintained at 25$\pm$10 nm. The resulting CL spectral map is shown in Fig. 2a. The position of individual CL peaks shows a pronounced spectral shift, closely following the change of the local fibre radius (Fig. 2c) and not with the distance from the apex, as would have been expected for interference between waves traveling in the fibre axis direction. For quantifying the change in the energy spacing, we calculated the Fourier transform of the CL spectra, shown in Fig. 2b (top) and compared it to the Fourier transform of simulated electron energy loss spectra (bottom), as discussed in the following section. The lowest order Fourier peak, signifying the cavity round-trip time $T$, changes from 115 to 145 fs ($\pm$5 fs) within the measurement region. The corresponding energy spacing $\Delta E=h/T$ (extracted from the position of the first Fourier peak maximum in Fig. 2b) is plotted in Fig. 2d (red) and compared to the energy spacing resulting from the ray-picture resonator equation (blue), as discussed above, and the energy spacing based on simulated EEL spectra (black). The difference between the experimentally and theoretically achieved energy spacings may be causes by an offset in the optical measurement of the fibre radius.       
	
	The Fourier-transformed experimental spectra are in good qualitative agreement with the simulated loss probability. The relative amplitude of high-harmonic peaks strongly depends on the damping time constant of the modes, experimentally affected by the taper angle and the surface quality of the silica fibre. In addition, the phase of the Fourier transform (not shown) is close to zero, as expected from the linear scaling of resonance energies with the azimuthal order $\nu$.
	
	Due to the spatial extent of the evanescent field of both the whispering-gallery modes and the moving electron, the electron-light coupling strength depends on the distance $b$ between the electron and the fibre surface. In Fig. 2e, the relative CL intensity within the photon energy range of 2.25 eV $<$ E $<$ 2.8 eV and the relative intensity of simulated EEL spectra are plotted as a function of $b$, showing an exponential dependency with an experimental decay length of 54 nm (simulation: 47 nm).

	\begin{figure*}
		\includegraphics[scale=0.8]{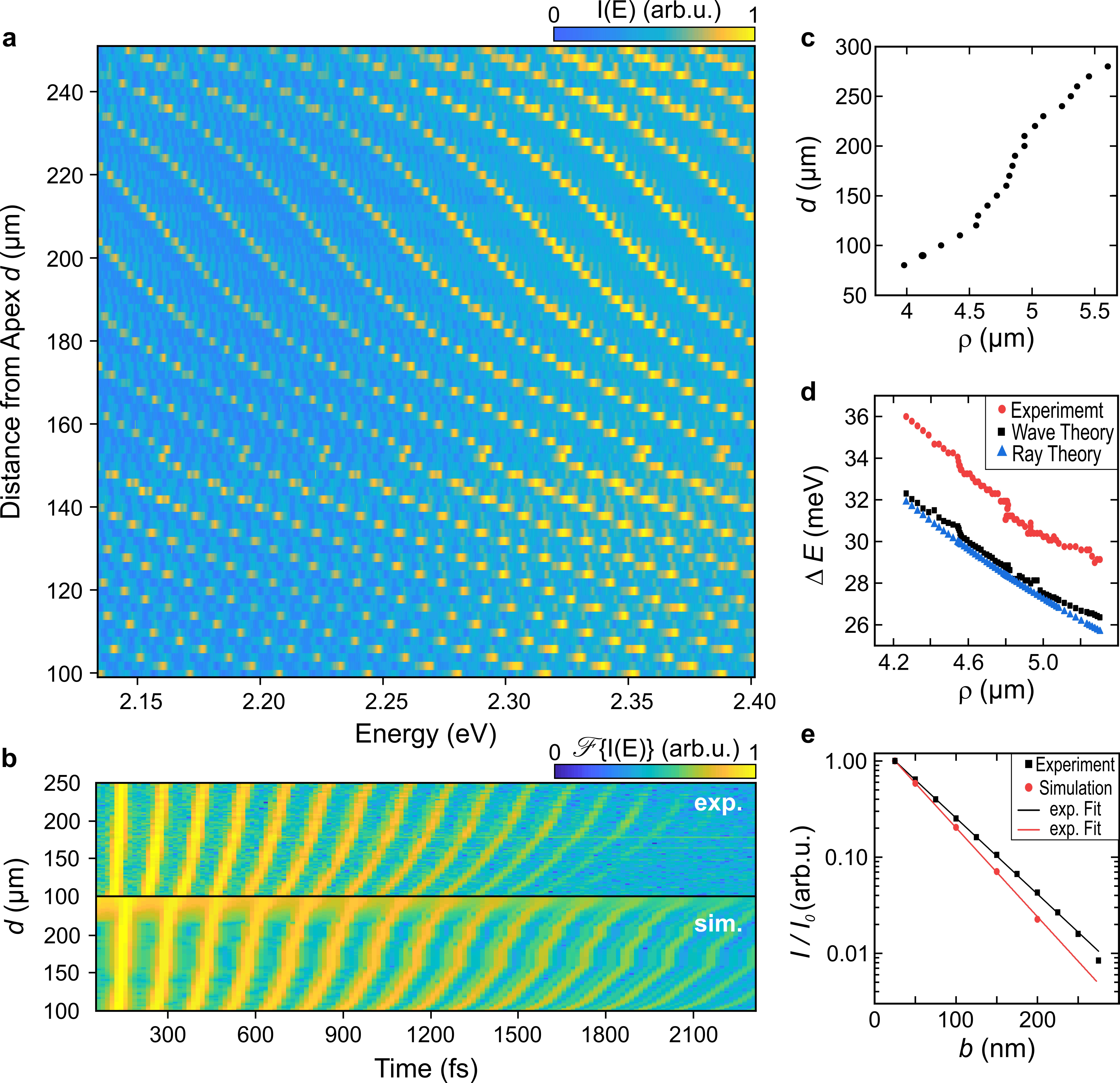}
		\caption{Dependence of coherent cathodoluminescence spectra on local fibre geometry. \textbf{(a)} CL spectral map recorded for different distances $d$ of the electron beam from the fibre apex. The WGM resonance peaks are shifting towards lower energies with increasing distances from the apex, following the increase of the local fibre diameter (impact parameter $b\approx25$~nm). \textbf{(b)} Fourier transform of the experimental CL spectra (top) and simulated electron energy loss probability (bottom), demonstrating the change in round-trip time with changing fibre radius. The large number of higher harmonics demonstrates the high quality of the tip-shaped resonator. \textbf{(c)} Local fibre radius $\rho$ determined from optical microscopy images at distances $d$ from the fibre apex. \textbf{(d)} Comb spacing $\Delta E$ plotted over the local fibre radius $\rho$. The energy spacing decreases with increasing radius due to larger round trip times $T$. \textbf{(e)} Dependence of the experimental CL intensity and the simulated EEL probability on the impact parameter $b$, demonstrating an exponentially decreasing coupling strength.}
		\label{figure2}
	\end{figure*}

	To further elucidate the coupling of fast electrons with whispering-gallery modes, we considered a theoretical model taking into account a cylindrical dielectric medium and the retarded back-action of the induced polarization on the passing electrons (see Methods and Refs. \cite{ochiai2004relativistic,talebi2015excitation,yalunin2016theory}). In the following, we exemplarily discuss the resulting simulated EEL probability spectrum $\Gamma (E)$ for a dielectric cylinder with radius $\rho=$ \SI{4.5} {\micro\metre}. Optical modes in the cylinder are characterized by the axial momentum $\beta$ (along the fibre axis) and the azimuthal and radial mode orders $\nu$ and $l$ \cite{matsko2006optical}. The momentum-resolved EEL probability map (Fig. 3a), shows a sequence of parabolic dispersion relation with increasing cut-off frequencies for a set of azimuthal orders. A weaker sequence of parabolic dispersion curves is related to WGM with higher radial order, which exhibit a less intense evanescent field leaking outside of the fibre, resulting in a weaker coupling with the electron beam. An unambiguous assignment of the individual features in the EEL map is facilitated by the momentum-integrated, mode-specific scattering probability, shown in Fig. 3b. Summing the scattering probability over all azimuthal modes, results in a EEL spectrum consisting of a series of approximately equidistant peaks (Fig. 3c), closely following the comb-like experimental CL spectra (cf. Fig. 1a). Each peak represents the energy cutoff of an individual azimuthal mode order, i.e. the minima of the dispersion parabolas ($\beta=0$) in Fig. 3a. The broadening of the resonance peaks towards higher loss energies results from the parabolic dispersion relation. Furthermore, we note that the simulated spectral density is in the range of $10^{-4}$/eV, similar to the CL probability densities observed experimentally (cf. Fig. 1c). The additional spectral substructure observed in the experiment may be related to further wave resonances due to the slightly conical shape of the resonator not considered in the simplified model applied here.

	\begin{figure}
		\includegraphics[scale=1]{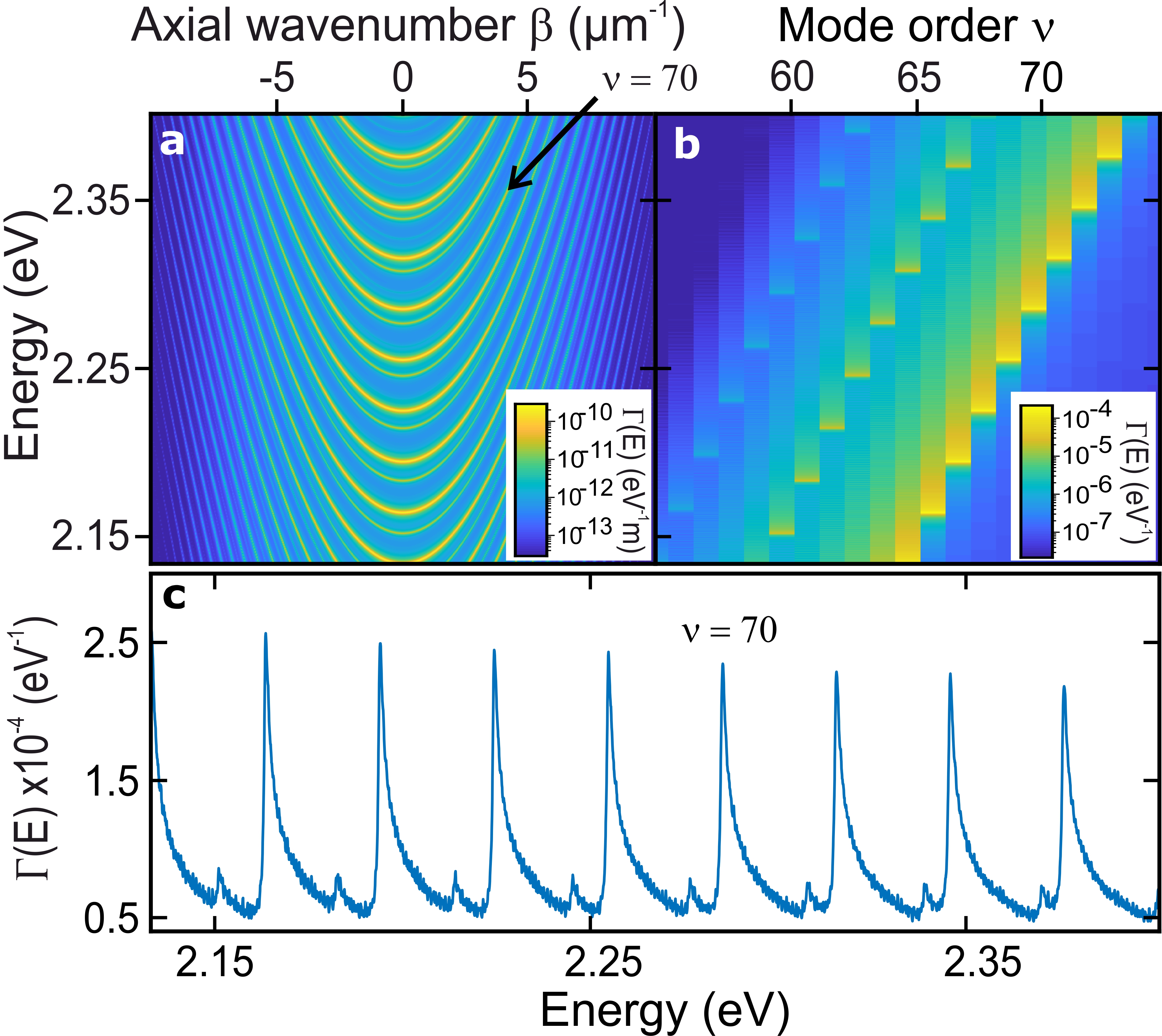}
		\caption{Simulated electron energy loss probability near a high-Q cylindrical silica resonator. \textbf{(a,b)} Loss probability $\Gamma(E)$ resolved in the axial wavenumber $\beta$ (a) and the azimuthal mode order $\nu$ (b). \textbf{(c)} Total loss probability spectrum obtained by integration over all azimuthal modes orders and axial wavenumber. The comb-like structure results from the minima in the dispersion curves, shown in (a). (fibre radius $\rho=$ \SI{4.5}{\micro\metre})}
		\label{figure3}
	\end{figure}
	
	The quality factor (Q-factor) of a resonator is a measure for the lifetime of a pulse within the cavity in units of the resonator round-trip time and thus represents the ratio between the stored energy and the dissipated energy per pulse round-trip \cite{vahala2003optical}. We note, that for the simulated spectra shown in Fig. 3c, we phenomenologically incorporated a finite excitation lifetime, by considering a refractive index of the dielectric medium with a small (0.02\%) imaginary component. 
	The Q-factor can be extracted from the experimental resonance spectra by considering the ratio between the resonant peak energy $E_0$ and its linewidth $\Delta E_{\mathrm{FWHM}}$ \cite{kavokin2017microcavities}. Fig. 4 shows a series of CL spectra with different Q-factors for selected distances $d$ from the fibre apex. The sharpest resonance features and thus the highest Q-factor ($Q\approx 700$) are obtained in the region around $d$ = \SI{200} {\micro\metre}. Following Fig.~2a,c, this fibre region shows the smallest $d$-dependence of the resonant peak position, corresponding to the fibre section with a close-to-cylindrical fibre geometry. This finding suggests that the currently achieved experimental Q-factors are still dominated by the macroscopic taper angle of the fibre tip and not by intrinsic loss mechanisms, such as inhomogeneous surfaces, defects or mode leakage.

	\begin{figure}
		\includegraphics[scale=1]{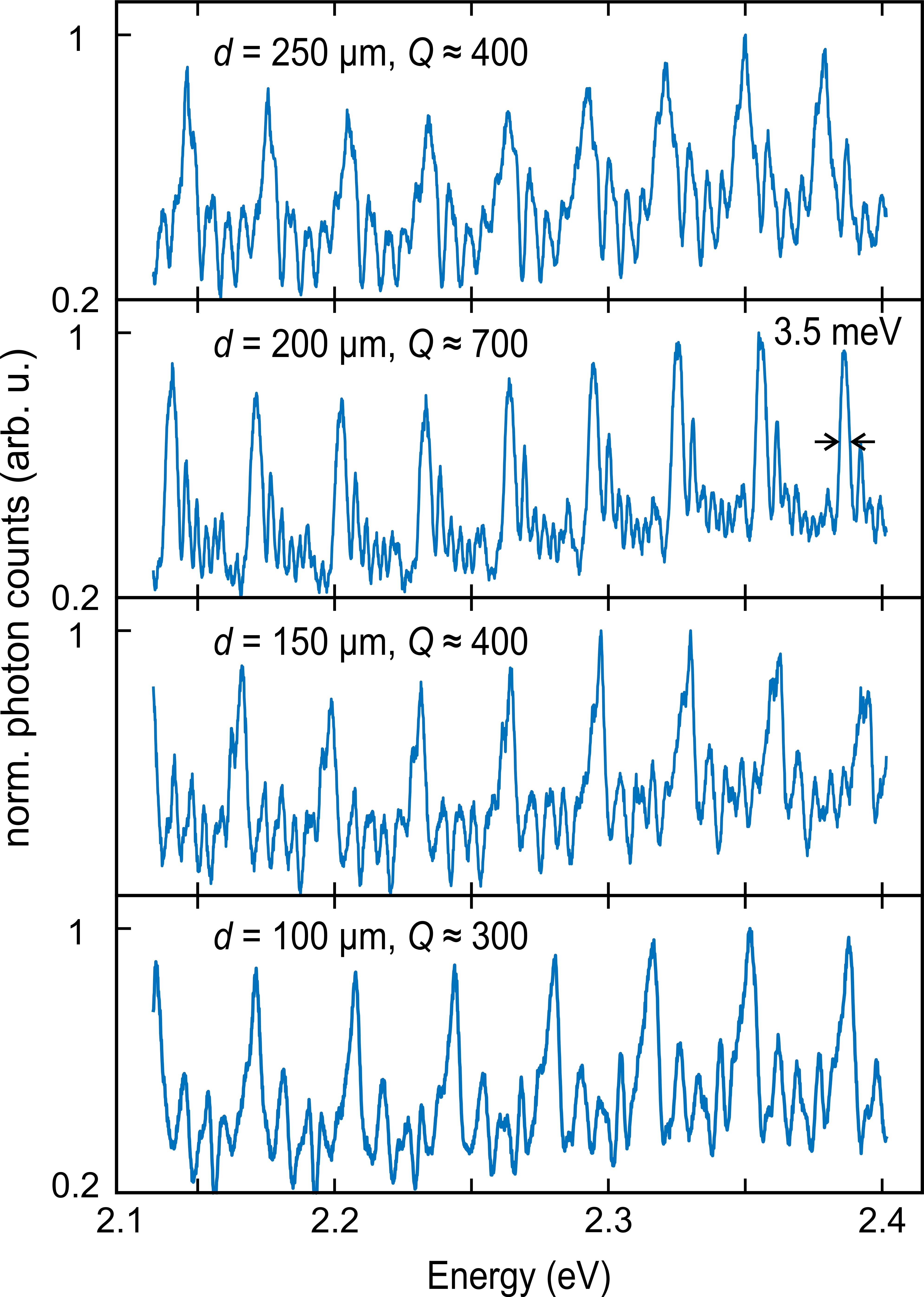}
		\caption{WGM quality factor. High-resolution CL spectra for different section of the fibre tip show a change in the peak width of the resonant features. Largest Q-factor of $Q\approx 700 \pm 50$ is found in the region around $d$ = \SI{200} {\micro\metre} where the local taper angle of the fibre tip is smallest (cf. Fig. 2c).}
		\label{figure4}
	\end{figure}


	For the non-optimized resonator geometry studied here, the highest observed Q-factor already results in a light storage time of approximately 500 fs. For a 200-keV electron beam with a current of 10 nA, the average temporal spacing between adjacent electrons in the beam is about 16 ps, so that resonator-mediated electron-electron correlations might be achievable by utilizing improved resonator geometrys \cite{collot1993very,gorodetsky1996ultimate}.

	In contrary to plasmonic excitations of metallic nanostructures which undergo rapid dephasing \cite{klar1998surface}, electron induced WGMs in dielectric microresonators therefore appear to be a promising tool for achieving mediated electron-electron and electron-photon entanglement and for studying electron energy loss probabilities scaling non-linearly with the electron beam current \cite{kfir2019entanglements}.
	Naturally, to this end, the coupling efficiency between the passing electron and the light mode needs to be optimized. Potential strategies include increased interaction times (narrowing the CL bandwith), achieved, for example, by larger taper diameters or in velocity-matched stadion-type resonators \cite{kfir2019entanglements}, or by utilizing the interaction with low-energy electron pulses in strongly confined fields or phase-matched media \cite{gulde2014ultrafast,talebi2019interference}. Finally, it should be mentioned that photon counting statistics and correlation measurements may provide further insights into the coupling between the electrons as fermionic particles and the bosonic optical excitations in the resonator.\\   
	In summary, we have experimentally demonstrated and studied the generation of comb-like broadband cathodoluminescence spectra by relativistic electrons passing a silica fibre taper. The experimental results are in agreement with numerical simulations of the corresponding EEL spectra. Our findings show that the Q-factor of the electron-induced whispering-gallery modes strongly depends on the resonator geometry, so that CL light storage times in the nanosecond to microsecond regime are within reach.

	\section*{Methods}

	\subsection{TEM Setup.}
	Microfibre tips with a tapered profile were extruded at the end of a multimode optical silica fibre (\SI{200}{\micro\metre} core diameter, polymer cladding with \SI{12.5}{\micro\metre} thickness, 0.5 numerical aperture) by a CO$_2$-laser heating procedure. The tapered fibre region with diameters below \SI{10}{\micro\metre} extended across a length of approximately \SI{200}{\micro\metre}, with a typical taper half-angle of approximately  \SI{0.3}{\degree} (see Fig.1). The polymer cladding is removed within the tapered section of the fibre. 
	The investigation of coherent cathodoluminescence from the fibre was performed in a transmission electron microscope (JEOL JEM2100F) operated at 200-keV electron energy. 
	Using a custom-built TEM holder, the microfibre tips were introduced into the microscope pole piece gap, with the non-tapered end of the fibre connected to a low noise-spectrometer (Kymera 328i with iDus 416, spectral resolution $\Delta\lambda\approx 0.5$~nm). Cathodoluminescence spectra were acquired with an exposure time of 15 s within a photon energy range from 1.4 eV to 3 eV, positioning the electron focal spot (10-nm spot diameter, ~680 pA beam current) with an impact parameter of $b\approx$25 nm relative to the fibre surface and at different distances $d$ from the fibre apex. For extracting absolute photon count rates, we considered the quantum efficiency of the detector and the grating efficiency as provided by the spectrometer manufacturer. In addition, due to the numerical aperture of the employed multimode fibre, we estimate that less than 20 \% of the generated CL light is guided along the fibre to the entrance of the spectrometer. 
	Depending on the electron beam current and the free-standing length of the fibre, moderate charging effects are visible. Electron beam alignment was performed in the presence of an approximately constant electrostatic field of the charged fibre.

	\subsection{Numerical Model.}
	The CL emission at the tapered microfibre was modeled considering the electron energy loss probability rate $\Gamma_{\mathrm{EEL}}(\omega)$ of a fast electron passing an infinitely long silica cylinder of a diameter corresponding to the experimentally probed tip section. The loss probability rate can be obtained \cite{de2010optical} from the Fourier components of the induced electric field $\mathbf{E}^{\mathrm{ind}}\left[\mathbf{r}_{e}(t), \omega\right]$ in the sample along the electron trajectory  $\mathbf{r}_{e}(t)$ with corresponding velocity vector $\mathbf{v}$, yielding
	
	\begin{equation}
	\Gamma_{\mathrm{EEL}}(\omega)=\frac{e}{\pi \hbar \omega} \int d t \operatorname{Re}\left\{\mathrm{e}^{-\mathrm{i} \omega t} \mathbf{v} \cdot \mathbf{E}^{\mathrm{ind}}\left[\mathbf{r}_{e}(t), \omega\right]\right\}.
	\end{equation}
	For a 200-keV electron passing a cylinder of given diameter and permittivity, the energy loss probability was calculated by means of a transition-matrix method recently presented by Yalunin \textit{et al.} \cite{yalunin2016theory}. Here, the externally induced electric field in the cylinder as well as the field of the fast electron are decomposed in terms of cylindrical vector harmonics containing the axial wavenumber $\beta$ and an angular and radial dependency of the fields according to the corresponding mode orders $\nu$ and $l$. The expansion coefficients of the cylindrical vector harmonics, given by $ w_{\beta \nu}$  and $\overline{w}_{\beta \nu}$ for the field of the fast electron and by $a_{\beta \nu}$ and $\overline{a}_{\beta \nu}$ for the induced electric field, are connected by the transition matrix $T_{\beta\nu}$ according to
	\begin{equation}
	\left(\begin{array}{l}{a_{\beta \nu}} \\ {\overline{a}_{\beta \nu}}\end{array}\right)=T_{\beta \nu}\left(\begin{array}{c}{w_{\beta \nu}} \\ {\overline{w}_{\beta \nu}}\end{array}\right).
	\end{equation}
	$T_{\beta\nu}$ depends on the diameter of the silica cylinder and its frequency-dependent permittivity $\epsilon(\omega)$ . For the numerical simulations discussed here we utilized the Sellmeier equations for the permittivity, with corresponding coefficients for fused silica given by Ref. \cite{malitson1965interspecimen}.  
	The energy loss probability in this formalism is obtained as
	\begin{equation}
	\Gamma(\omega)=\frac{e^{2} }{2 \pi^{2} k^{2} \epsilon_{o} \hbar} \sum_{\nu=-\infty}^{\infty} \int_{-\infty}^{\infty} \operatorname{Re}\left(a_{\beta \nu} w_{\beta \nu}+\overline{a}_{\beta \nu} \overline{w}_{\beta \nu}\right) d \beta,
	\end{equation}
	and consists of the contributions by the induced field of each mode with angular mode order $\nu$, integrated over all axial wavenumbers $\beta$.

	\bibliography{references}

	\begin{acknowledgement}
	We acknowledge financial support by the Volkswagen Foundation as part of the Lichtenberg Professorship "Ultrafast nanoscale dynamics probed by time-resolved electron imaging" and the DFG with the priority program 1840 "Quantum Dynamics in Tailored Intense Fields". We thank the group of N. Nilius from the University of Oldenburg for kindly providing the CO$_2$ laser setup for fibre tapering. The electron microscopy service unit of the University of Oldenburg is acknowledged for their support in using the TEM.
	\end{acknowledgement}

\end{document}